\documentclass[aps,preprint,epsfig,rotate]{revtex4}
\usepackage{graphicx}
\usepackage{bm}

\def\ie{i.e.\ }

\def\PD#1#2{\frac{\partial #1}{\partial #2}}    
\def\mtr{\mathrm{Tr}}
      \def\hh2{$\mathrm{H+H_2}$}
      \def\h2{$\mathrm{H_2}$}
      \def\t2{$\mathrm{T_2}$}

      \def\dd2{$\mathrm{D+D_2}$}
      \def\tt2{$\mathrm{T+T_2}$}
\def\mrm #1{\mathrm{#1}}
\def\k #1{|#1\rangle}
\def\b #1{\langle #1|}
\def\sfl {{\sf{L}}}
\input epsf
\begin{document}

\title{Decoherence Effects in Reactive Scattering}
\author{Heekyung Han and Paul Brumer}
 \affiliation{ Chemical Physics Theory Group, Department of Chemistry,\\
 and Center for Quantum Information and Quantum Control,
\\University of Toronto\\ Toronto, Canada  M5S 3H6}

\date{\today}

\begin{abstract}
Decoherence effects on quantum  and  classical dynamics
in reactive scattering  are examined using a Caldeira-Leggett type
model.  Through  a
study of dynamics of the collinear   \hh2 reaction  and the transmission
over simple one-dimensional barrier potentials,  we  show that decoherence
leads to improved agreement between quantum and classical reaction and
transmission probabilities, primarily by increasing the energy dispersion
in a well defined way. Increased potential nonlinearity is seen to require
larger decoherence in order to attain comparable quantum-classical agreement.
\end{abstract}
\maketitle

\vspace{0.2in} \noindent

\pagebreak

\section{Introduction}\label{formalism}

Isolated molecular collisions have been the subject of theoretical study
over the past half century. Amongst the successful tools in the field are
classical trajectory approaches, whose justification often relies upon
the relatively small deBroglie wavelengths in a chemical reaction.
However. most chemical reactions occur in the condensed phase. Such an
environment, due to its randomness, tends to decohere\cite{decoherence_review}
a quantum system, driving it
towards the classical limit. Hence, we would anticipate that the utility
of classical methods would gain further validity from this loss of phase
information. In this paper we investigate the extent to which quantum
effects are lost, in chemical reactions, due to the environment.
Specifically, we examine, using a Caldeira-Leggett type of approach
\cite{joos,DPE,unruh} applied to the paradigmatic collinear H + H$_2$
system \cite{hh2research}, the effects of the environment on the two
predominant quantum effects in reactive scattering: tunneling and
resonances. We find, for reasonable values of the decoherence, that the
quantum dynamics of the (system + environment) approaches that of the
classical dynamics of the (system + environment). A detailed analysis of
the tunneling and resonance regimes is also provided through a study of
simple one dimensional models. Improved decoherence-induced
quantum-classical agreement in the reactive cross section is found, in
both the tunneling and resonance regions, to stem from increasing energy
dispersion with increasing decoherence strength and increasing interaction 
time.

The   general issue  of  the transition  from  quantum  to  classical
dynamics remains a subject of considerable interest and activity.
Traditionally, classicality is argued to  emerge as $\hbar$ goes to zero
or equivalently as the system mass goes to infinity. A comprehensive
Liouville-based approach  \cite{Wilkie} provides a more complete
description of this limiting process for both integrable and chaotic
systems.  Alternatively, decoherence, \ie, the loss of quantum coherence
as induced by the interaction of a system with its environment, has been
proposed to be an essential ingredient for the utility of classical
mechanics in the real world  \cite{decoherence_review}.  Early theoretical
studies on model systems successfully demonstrated that  decoherence is effective in
bringing about classical behavior in the macroscopic  limit \cite{zurek}.
In addition, our earlier study\cite{gong} extended this work to Hamiltonian systems, 
showing the emergence of classical behavior due to decoherence in a bound
Hamiltonian system.

Despite numerous papers, and some applications in chemistry
\cite{rossky,prezhdo} little work has been done on the role of
decoherence in inducing classicality in scattering problems, such as
chemical reactions.   We do so below, via a simple master equation for
the reduced density matrix $\rho$, i.e. the density matrix for the
scattering system after averaging over the external bath. The essential
observable under consideration is the reactive scattering cross section,
which displays both quantum tunneling and resonance scattering.

\section{Decoherence: Formulation and Computation}

Consider a scattering event that occurs in the presence of an external
environment, focusing attention on the short time regime wherein
decoherence, but not energy transfer between the system and environment,
occurs. A master equation for $\rho$ for such a setup has been obtained
by Joos and Zeh \cite{joos} for an isotropic environment that elastically
scatters from the system of interest with negligible momentum transfer.
With the additional assumption that the $\langle x|\rho|x'\rangle$ of
interest are those with $k|x-x'| << 1$, where $k$ is the typical
wavenumber of the environment and $x$ is the system coordinate, 
they obtained an equation for $\rho$ that
is the same as the oft-cited model of Caldeira and Leggett \cite{DPE}
and that of Unruh and Zurek \cite{unruh}, when  dissipation is ignored.
The latter corresponds to ignoring the transfer of momentum to the
environment, a reasonable assumption for the short time scales associated
with the collision process.

For a system with two degrees of freedom, the resultant quantum master
equation is given as \cite{DPE,gallis}:
\begin{eqnarray}
\frac{\partial \rho}{\partial
t}=\frac{1}{i\hbar}[H,\rho]-\frac{D}{{\hbar}^2}[x,[x,\rho]] -
\frac{D}{{\hbar}^2}[y,[y,\rho]] \label{operator}
\end{eqnarray}
with  the system Hamiltonian $H$  and the decoherence constant $D$.  To
compare to the classical result it is convenient to consider this
equation in the Wigner representation \cite{DPE}:
\begin{eqnarray}
\frac{\partial }{\partial t}\rho^{W} & =
&\sfl_{\mrm{cl}}\rho^{W}+\sfl_{\mrm{q}}\rho^{W}+\sfl_{\mrm{D}}\rho^{W};\label{deco} \\
\sfl_{\mrm{cl}}&\equiv&\{H, \}_{\mrm{PB}},\label{pb}\\
\sfl_{\mrm{q}}&\equiv&\sum_{(l_{1}+l_{2})=\ odd} \frac{1}{l_{1}!\
l_{2}!}\left(\frac{\hbar}{2i}\right )^{(l_{1}+l_{2}-1)}
\frac{\partial^{(l_{1}+l_{2})}V(x,y)}{\partial x^{l_{1}}\partial
y^{l_{2}}} \frac{\partial^{(l_{1}+l_{2})}} {\partial
p_{x}^{l_{1}}\partial p_{y}^{l_{2}}},\label{qmcont}\\
\sfl_{\mrm{D}}&\equiv& D\left(\frac{\partial^{2}}{\partial
p_{x}^{2}}+ \frac{\partial^{2}}{\partial
p_{y}^{2}}\right).\label{diffcont}
\end{eqnarray}
where $\{\cdot\}_{\mrm{PB}}$  denotes the classical Poisson bracket and
where $\rho^W \equiv \rho^W(p_x,p_y,x,y)$ denotes the Wigner transform of
$\rho$. Here, The $\sfl_{\mrm{cl}}$ generates the classical dynamics of
the closed system, and $\sfl_{\mrm{q}}$ generates the quantum corrections
to this dynamics, whereas $\sfl_{\mrm{D}}$ induces decoherence in the open
system. When $D=0$ Eq. (\ref{deco}) becomes the
quantum Liouville equation for the closed system.

Equation (\ref{deco}) is solved using the quantum state diffusion (QSD)
approach  \cite{gong, perci1} wherein one solves an associated stochastic
Schrodinger equation for the state vector given by
\begin{eqnarray}
\k{d\psi}=\left(-\frac{i}{\hbar}H -\sum_{m}\left[L_{m} -\langle
L_{m}\rangle_{\k\psi}\right]^2 \right)  \k\psi dt
+\sum_{m}\left[L_{m} -\langle L_{m}\rangle_{\k\psi}\right]\k\psi
dW_m.\label{QSD}
\end{eqnarray}
Here   the self-adjoint operators $L_{m}$ represent the coupling between
the system and environment, $ \langle L_{m}\rangle_{|\psi\rangle}=\b
{\psi} L_{m} \k \psi /\langle \psi \k \psi$, and $W_m$ is a complex
Wiener process  \cite{perci1}. Our specific implementation of the QSD
method takes the operators $L_m$ as $L_{1}$=$\frac{\sqrt{D}}{\hbar}\hat{
x}$ and $L_{2}= \frac{\sqrt{D}}{\hbar}\hat{y}$. Computationally, the
unitary component of Eq.  (\ref{QSD}) is integrated by the fast Fourier
transform method plus the split-operator  \cite{fleck}, as in a closed
quantum system. The terms involving $L_m$ are integrated using a second
order scheme \cite{secondorder}. The dynamics resulting from the
Stochastic Schr\"{o}dinger equation (\ref{QSD}), when averaged over many
realizations of the Wiener process, provides the solution to Eq.
(\ref{deco})  \cite{perci1}. Hence, the quantum reaction
probabilities for the open system are obtained by averaging over reaction
probabilities for each wavefunction.

A comparison of these results with the classical result means comparing
the quantum result of the (system + environment) with the classical
result for the (system + environment). The latter is obtained by  setting
$\sfl_{\mrm{q}}=0$ in Eq. (\ref{deco}), leading to the classical
Fokker-Planck equation for the classical density $\rho^{\mrm{cl}} \equiv
\rho^{\mrm{cl}}(p_x,p_y,x,y)$:
\begin{eqnarray}
\frac{\partial }{\partial t}\rho^{\mrm{cl}}  =
\sfl_{\mrm{cl}}\rho^{\mrm{cl}}+\sfl_{\mrm{D}}\rho^{\mrm{cl}}.
\label{fokkerp}\end{eqnarray} The resultant dynamics is equivalent to the
Langevin-It$\mrm{\hat{o}}$ equations with  Gaussian white noise for each
degree of freedom, which are solved by combining Monte Carlo sampling
from the initial distribution and  fourth-order Runge-Kutta integration.
Results in the absence of decoherence ($D=0$) were obtained directly by
solving  Hamilton's equations for the classical trajectories. The
classical reaction probabilities for both the closed and open cases are
obtained by counting the number of trajectories that are reactive.

\section{H + H$_2$}

\subsection{Computational Features}

We present results for collinear \hh2. Isotopic variants of this
system were also studied \cite{thesis}, with similar results to those
reported here. The  Hamiltonian for the collinear \hh2 system is
$H=\frac{1}{2\mu}(p_{x}^{2}+p_{y}^{2})+ V(x,y)$, where
$x$ and $y$  are the  mass-scaled Jacobi coordinates
of $R$ (the distance of $\mrm H$ from the center of mass of $\mrm{H_2}$)
and $r$ (the $\mrm{H_2}$ internuclear distance), respectively, and
$p_{x}$ and $p_{y}$ are the corresponding momentum operators. The reduced
mass $\mu$ $=\sqrt{[{m_1}{m_2}{m_3}/{(m_1+m_2+m_3)}]}$ where $m_i$ is the
mass of H. The Liu-Siegbahn-Truhlar-Horowitz potential energy
surface\cite{LSTH} was used for $V(x,y)$.

The  initial  wavefunction, placed far from the interaction region, can be
written as the product of a minimum uncertainty Gaussian wave packet
$F(x)$ describing the relative initial translational motion times the vibrational
eigenfunction
$\phi_{v}(y)$  for  $\mathrm{H_2}$   with vibrational quantum number
 $v$. That is,
    $\psi(x,y,t=0)=F(x)\phi_{v}(y)$,
with
   \begin{eqnarray}
   F(x)={(2\pi\gamma^2)}^{-1/4}\exp{[-\frac{{(x-x_0)}^2}{4\gamma^2}-\frac{ip_{x0}x}{\hbar}]}.
  \label{fr}
  \end{eqnarray}
Here $\gamma$ is the width of $F(x)$, and $x_0$ and $p_{x0}$
describe the locations of the initial wavepacket in position
and momentum, respectively. A
large value  of $\gamma$ results in a
spatially delocalized wavepacket that is localized in momentum
space.  The initial total energy, $E$, is a sum of the average
kinetic energy of the translational wavepacket and the vibrational
energy. In the study below we consider the ground vibrational
state ($v=0$) and vary the translational energy.
Note that the energy width $\delta E \equiv
\sqrt{\langle {H^2}\rangle -{\langle H \rangle}^2 }$ of the
initial wavepacket, of considerable interest later below, is
determined by the translational component. Thus, by choosing
$\gamma$ appropriately the wavepacket can be focused on one average
energy component of interest, $E$, with a narrow energy width.
Specifically, since the initial wavepacket is in the asymptotic
region ($x\rightarrow \infty$) and  $p_{x0}$ is chosen as
$p_{x0}=-\sqrt{2\mu (E-E_v)}$, $\delta E $ is given by $ \delta E
= \sqrt{ \frac {\hbar^4}{32 {\gamma}^4\mu^2}+ \frac {\hbar^2
(E-E_v)}{2 {\gamma}^2 \mu} }$,
 with vibrational energy $E_v$.
  Since the smaller the $\gamma$, the faster the numerical computation,
 $\gamma$  is taken as small as possible,  limited by the need to obtain an
 adequate  $\delta  E$ that depends on the resolution of
energy region studied. Our choice of $\gamma$=5.9 a.u. gives a
$\delta E\approx$ 0.068 eV for the initial wavepacket, which is 
somewhat larger, due to computational limitations, than that expected
analytically.

To examine quantum-classical correspondence requires an appropriate
comparison with an initial classical ensemble corresponding to the
above initial quantum state. For the translational component,
this is given by the Wigner function
corresponding to $F(x)$ in Eq. (\ref{fr}). However, there are
several possible choices for the vibrational component, 
which can be used to
make  the initial quantum and classical conditions as similar as
possible { \cite{wyatt,kuppermann,wignercm}}. For example, since
the initial state is a vibrational eigenfunction, the
corresponding Wigner function does not vanish at the classical
vibrational turning points, so that some phase points have
initial classical energies that do not equal the quantized
value. The resultant
 $\delta E$ for the vibrational motion differ for the initial quantum and
 classical density functions;
 the vibrational energy width of the initial quantum state is
zero but that of the classical density is $\hbar\omega/2$.  For
the case of $\mathrm{H_2}$, this equals 0.27 eV. This width is much larger
than that of 
the quantum counterpart whose entire energy width, 0.068 eV, comes
from the translational part. This leads to classical reaction
probabilities that are insensitive to decoherence.

Our results show that for studies of reaction probabilities versus energy,
it is crucial to have the same initial $\delta E$ for both the quantum and
classical systems. Hence, each classical trajectory is chosen to have the
fixed quantum vibrational energy, guaranteeing the same $\delta E$ as that
of the quantum system. We assume that \h2 is described as a Morse
oscillator and obtain the initial vibrational variables as in Ref.
 \cite{kuppermann,millerr}.

Below, two different values of $D$ are examined. The
smaller of the two, $D=2.47\times 10^{-35}$  kg$\cdot$J/s,
corresponds to $10^{-2}$ to $10^{-1}$ of typical $D$ values 
in solution (a value estimated in Appendix). The larger
$D$ is 20 times the size of the smaller. Both these values of $D$
are in the range extracted for $D$ from a pseudo-realistic model
problem \cite{yossi}. For the open quantum system we averaged over
72 realizations for the smaller decoherence and over 288
realizations for the larger one. For both the closed and open
classical systems $10^4$ trajectories are used. Below, the notation QM and CM
denote the quantum mechanical and classical probabilities,
respectively, in the absence of decoherence. The symbols QMD and
CMD denote results in the presence of decoherence.

We focus below on the effect of decoherence on the  total  reaction
probability $P^{R}(E)$, which  can be written as
\begin{eqnarray}
P^{R}(E)=\int^{\infty}_{0}J^{R}(t,E) dt =\Delta t \sum_{n=0}^{\infty} J^{R}(n\Delta t ,E),
\label{pre}
\end{eqnarray} 
with a chosen time increment  $\Delta t$.
Here the reaction flux, $J^{R}(t,E)$ is given as:
\begin{eqnarray}
J^{R}(t,E)=(\hbar/\mu)\mathrm{Im}\mathit{[\langle \psi_E(x,y=y_{\mathrm{I}},t)|\partial
\psi_E(x,y=y_{\mathrm{I}},t)/\partial y \rangle]},
\label{jrt}
\end{eqnarray}
where $\psi_E(x,y,t)$ is the system wavefunction at time $t$ that
evolves from an initial wavepacket
$\psi_E(x,y,t=0)$ that is very narrowly focused on $E$, and
the brackets denote the integration over $x$,
obtained  by  adding  all  the discrete contributions over the
entire range  of  $x$  along  a  dividing line $y=y_{\mathrm{I}}$
in the product region. 

The simulation uses  absorbing boundary
conditions that damp the wave packet with a sine masking
function near the edge of the grid in both reactant and product
channels  \cite{maskf}. We used the following parameters for QM
and QMD; the grid starts at $(x,y) = (x_\mrm{min},y_\mrm{min})$ =
(0.90,0.38), the mesh size $(\Delta x,\Delta y)$= (0.15,0.12), the
initial position $(x_0,y_0)$=(30, 1.35), and $y_{\mathrm{I}}=3.02$
in atomic units. For QM,  $\Delta t$= 0.14 fs,  and for QMD
$\Delta t$ is 0.073 fs for the smaller $D$ and 0.012 fs for the
larger $D$. The total time steps are chosen appropriately, keeping
$t_f =287$ fs constant. The classical reaction probability is
evaluated by counting the number of  trajectories that cross the
dividing line $y=y_{\mathrm{I}}^{\mathrm{CM}}$= 5 a.u.
 before the final time $t_f^{\mrm{CM}}$=290 fs. Numerical checks
of quantum and classical calculations simulations at
different temporal and spatial resolutions were carried out,
and  results for the closed systems were checked against 
previously computations\cite{wyatt,kuppermann,wignercm}.

\subsection{Computational Results}

Figures \ref{fig1m} and \ref{fig1m2}  display the classical   and quantum
reaction probabilities  as a function of  $E$ for
the \hh2 exchange reaction in the closed system
and open systems for both values
of $D$. For the closed system the results agree well with those of the
previous studies  \cite{kuppermann} and one sees large deviations between the
classical and quantum probabilities. In particular, note the strong
quantum resonance dips at $E \sim$ 0.9 eV $E \sim
$ 1.2 eV as well as the tunneling
that are all absent in the classical result. Upon introducing the
smaller of the decoherence, the resonances are damped out, as seen in
Fig. \ref{fig1m}, leading to better agreement between quantum and
classical reaction probabilities. Also notable is that the addition of
the decoherence in Fig. \ref{fig1m} enhances the reaction
probabilities at $E<$ 0.55 eV and suppresses them at 0.55 eV $<E<$ 0.7 eV,
for both the quantum and classical cases. This behavior is even more
pronounced in the case of the larger $D$, shown in Fig. \ref{fig1m2},
and is analyzed in 
Section \ref{tunnel} where tunneling in one dimension is examined in
detail. Note from Fig. \ref{fig1m} that the CMD and QMD results are still
in considerable disagreement. This disagreement disappears with the
higher $D$ [Fig. \ref{fig1m2}] where the classical and the quantum
results are in very good agreement over the entire energy region.
However, in this case most of the scattering structure is suppressed: the
reaction probabilities are elevated at energies below the threshold and
reduced above the threshold energy, and the $P_R$ curve is now only
slightly convex about $P_R = 0.5$.

Note that we also examined the case where the Wigner function is used for
the vibrational component of the initial classical ensemble. In this case the
classical reaction probabilities showed, in the threshold region, 
a slower rise with energy than the quantum
counterparts in both the closed and open systems. This behavior is due to
the larger $\delta E$ of the initial classical density. In this case
the resultant
insensitivity of the classical reaction probabilities to decoherence leads
to a persistent discrepancy between quantum and classical reaction
probabilities, especially   in the threshold region. Hence, in this case,
basing the classical density on the Wigner function provides misleading
results.

Having demonstrated that classical dynamics emerges in the presence of
decoherence for the H + H$_2$ reaction, it remains necessary to identify
the specific mechanism by which this occurs in the Caldeira-Leggett 
type model.
To do so, we first show below that the system energy dispersion $\delta E =
[\langle H^2 \rangle - \langle H \rangle^2]^{1/2}$ increases in these models
as $\sqrt{D}/m$. Tunneling and
resonance behavior for a simpler one dimensional model 
are then examined to establish the role of this
increasing $\delta E$ as the source of the decoherence-induced
quantum-classical correspondence. We then return to remark on the case of
reactive H+H$_2$.

\section{Time Evolution of the Energy Dispersion $\delta E$}

The friction free version of the Caldeira-Leggett master equation [Eq.
({\ref{operator})] is known to display an increase in system
energy due to the $D$ dependent
decoherence term (the so-called environmental localization term). 
This energy increase would be compensated for,
at far longer times, by the transfer of energy back to the
environment \cite{gallis}. More significantly, for our purposes,
is the fact, shown below, that over similar timescales 
the square of the energy dispersion $(\delta E)^2
= \langle H^2 \rangle - \langle H \rangle^2$ increases as $D/m^2$,
where $m$ is the system mass. To see this, we extend the method of
Ref. \cite{gallis} to the case of energy dispersion.

For simplicity, consider a one dimensional case, in which Eq.
({\ref{operator}) reduces to
\begin{eqnarray}
\frac{\partial \rho}{\partial
t}=\frac{1}{i\hbar}[H,\rho]-\frac{D}{{\hbar}^2}[x,[x,\rho]] .
\label{effectivem}\end{eqnarray}
For any observable $O$ that does not have explicit time dependence:
 \begin{eqnarray}
 \frac{d\langle O\rangle}{dt}&=&\frac{d}{dt}\mathrm{Tr}
 ( O\rho ) \nonumber \\
    &=&\mathrm{Tr} {\big\lgroup} O\PD{\rho}{t}  {\big\rgroup} \nonumber \\
    &=&  \frac{1}{i\hbar}\mathrm{Tr} {\big\lgroup}O[H,\rho]
    {\big\rgroup}
    - \frac{D}{\hbar^2}\mathrm{Tr}{\big\lgroup}O[x,[x,\rho]]{\big\rgroup}.
   \label{operator2} \end{eqnarray}
We note two useful relations: the cyclic invariance of the trace, which
gives
 \begin{eqnarray}
\mathrm{Tr}(A[B,C])=\mathrm{Tr}([A,B]C), \label{cyclic}
\end{eqnarray}
and the identity
\begin{eqnarray}
[A,f(B)]=df/dB[A,B], \label{commutator}
\end{eqnarray}
which holds when $[A,[A,B]]=[B,[A,B]]=0$.
Using Eqs. (\ref{cyclic}) and (\ref{commutator})
one can verify the following relations: 
\begin{eqnarray}
\mtr(x[x,[x,\rho]])&=&0,  \nonumber \\
\mtr(V(x)[x,[x,\rho]])&=&0,  \nonumber \\
\mtr(V^2(x)[x,[x,\rho]])&=&0,  \nonumber \\
\mtr(p_x[x,[x,\rho]])&=&0,  \nonumber \\
\mtr(p_x^2[x,[x,\rho]])&=&-2\hbar^2,  \nonumber \\
\mtr(p_x^4[x,[x,\rho]])&=&-12\hbar^2\langle p_x^2\rangle,  \nonumber\\
\mtr{\big\lgroup}\left\{p_x^2
V(x)+V(x)p_x^2\right\}[x,[x,\rho]]{\big\rgroup}&=&-4\hbar^2\langle
V(x)\rangle.\label{relation}
\end{eqnarray}
which are used to evaluate the effect of the environmental localization
term.

Given these expressions the time-evolution of the energy
dispersion can be evaluated. Specifically:
\begin{eqnarray}
\frac{d(\delta E)^2}{dt}&=&\frac{d\langle H^2 \rangle}{dt}-2\langle
H\rangle\frac{d\langle H \rangle}{dt}\nonumber\\
&=&D\left( 3{\langle p_x^2 \rangle}{/m^2}+2{\langle
V(x)\rangle}{/m}\right) -2\left({\langle p_x^2\rangle}{/2m}+\langle
V(x)\rangle\right)D/m
\nonumber\\
&=&2D\langle  p_x^2\rangle/m^2,
\label{energywidth}
\end{eqnarray}
where $H$ is the system Hamiltonian given by $H=p_x^2/2m+V(x)$ and where we
have used the result \cite{gallis} that $d \langle H \rangle/dt = D/m$.
The extension to the two-dimensional H + H$_2$ system is straightforward,
giving

\begin{eqnarray}
\frac{d(\delta E)^2}{dt}&=&\frac{2D}{\mu ^2}{\langle p_x^2
\rangle}+\frac{2D}{\mu ^2}{\langle p_y^2 \rangle},
\label{delta_e2_2d}
\end{eqnarray}
and the rate of the average energy increase becomes $4D/\mu$.

\section{Decoherence in One Dimensional Scattering}

\subsection{Model Systems}

As is well known
\cite{model}, systems with increasing nonlinearity are expected to
show increasing quantum contributions to the dynamics. 
Hence we consider the effect of decoherence on
several one-dimensional barrier potentials of different
nonlinearities. Two
different types of barriers are designed for each degree of
nonlinearity, a single barrier potential (denoted SB) showing no
resonance and a double barrier potential (denoted DB) that shows a
typical resonance in the transmission probability versus initial
energy. The insights gained from these simple systems help shed
light on the \hh2 results.

For  the one-dimensional cases,   the computational formulae
introduced above reduce to the analogous equations for one
degree of freedom. The numerical methods used to obtain
transmission probabilities in the quantum and classical systems
are the same as described in the  \hh2 reaction case, the only
difference being that here the absorbing potential at
the grid boundary is not implemented.
Instead, a sufficiently large position range is
used so that the wavepacket does not encounter the boundary during
the time evolution. The labels
QM, CM, QMD and CMD have the same meaning here as described in the
\hh2 case. For the open quantum system a total of 500-1000
wavepackets are used, and for both the closed and open classical
systems $10^4$ trajectories are used. All variables below are in
dimensionless units.

The Hamiltonian for  one-dimensional scattering is
\begin{equation}
H= \frac{1}{2m}p_{x}^{2}+ V(x)=-\frac{\hbar^2}{2m}\frac{d^2}{dx^2} + V(x),
\label{h1}
\end{equation}
with mass $m$  and potential $V(x)$. Here $m=1$ and $\hbar=0.1$.
The initial wavefunction  is chosen as a minimum uncertainty
Gaussian wave packet $F(x)$ [Eq. (\ref{fr})], and the associated
Wigner function is used as the initial classical ensemble. In
order to set $\delta E$ of the initial wavepacket the same
method as described above is used, except that here $E_v$ is zero, and the
mass and $\hbar$ are different. We choose $p_{x0}=\langle
p_x\rangle=\sqrt{2mE-{\hbar}^{2}/4{\gamma}^{2}}$, and thus $\delta
E = \sqrt{ -\frac {\hbar^4}{32 {\gamma}^4 m^2}+ \frac{\hbar^2 E}{2
{\gamma}^2 m} }$.
Specifically, we take
\begin{eqnarray}
V_\mrm{SB}(x) &=& V_0[\cosh{ (\alpha x) }]^{-2},\\
V_\mrm{DB}(x) &=& V_\mrm{SB}(x-\beta)+V_\mrm{SB}(x+\beta)
\label{potentials}
\end{eqnarray}
as potentials for the SB case and for the DB case, respectively. The
degree of system nonlinearity is estimated via a characteristic potential
length, $ \chi_n(\equiv |{\partial_{x}V}/{\partial_{x}^{n+1}V}|^{1/n})$
 \cite{model}: the smaller the $\chi_n$, the larger the nonlinearity. Since
the derivatives  of these potentials obey the following relations:
\begin{eqnarray}
d^{2n+1}V(x)/dx^{2n+1}&=&(-4{\alpha}^2)^{n}dV(x)/dx, \\
d^{2n}V(x)/dx^{2n}&=&(-4{\alpha}^2)^{n-1}dV^2(x)/dx^2,
\label{dV2n}
\end{eqnarray}
$\chi_{2n+1} = 1/2\alpha$ for  both the  SB and DB cases. Hence,
increasing $\alpha$ leads to the larger nonlinearity. For the SB case, we
examine two cases, a weakly nonlinear potential (denoted SBW) and
a strongly nonlinear one (denoted SBS) with $\alpha=0.5$ and
$\alpha=10$, respectively. For the DB case, $\beta=0.1$ and 
only a strongly nonlinear case with $\alpha=10$ is examined. All $V_0$
are chosen to give the same barrier height of 2.0 and numerical
parameters are chosen to ensure accuracy \cite{numbers}.

Figure \ref{fig10} shows the potential energy for SBW, SBS, and DB along
with their first derivatives. Since the nonlinearity is inversely proportional to the width of the potential energy, a narrow/wide barrier has a large/small nonlinearity. This  can be clearly seen  in Fig. \ref{fig10}.  Thus, noting that the first derivatives of SBS and
DB are approximately  20 times larger than that of SBW,  we anticipate
that the quantum corrections for SBS and DB should be comparable to one
another and larger than that of SBW and that the magnitude of the
decoherence needed for the emergence of classicality in the SBW case
should be smaller than that of the SBS case. This is indeed observed
below.

\subsection{Tunneling Regime}
\label{tunnel}

Figure \ref{fig11_0} shows the results of the tunneling
calculations. For cases with no decoherence, in
addition to the classical (CM) and quantum (QM) results, we show the exact
quantum-mechanical transmission probabilities for the single
barrier potentials, denoted QMA. They are given by
\cite{liboff}
\begin{equation}
P_R={ {\sinh }^{2} { (\pi p_{x0}/\hbar \alpha) } }/
 { [ {\sinh}^{2}  {  (\pi p_{x0}/\hbar \alpha) } + {\cosh}^{2} { \{\pi/2
 \sqrt{c^2-1}\}}] }
 \end{equation}
where  $c^2  \equiv  {8mV_0} / {\hbar^2 \alpha^2} \: >\:1$, and used as
a computational check. In the absence of
decoherence the strongly nonlinear cases, SBS and DB, show larger
differences between the quantum and classical results than does
the weaker nonlinearity case. Further, tunneling in the quantum
cases is clear, as is the broad resonance around $E=2.6$ in the
double barrier case, reflecting the presence of a metastable state
in the quantum double barrier.

In Fig. \ref{fig11_0}, the decoherence shown is of magnitude that 
lead to roughly good agreement between quantum and classical
results. In panel (a) this corresponds to  $D=6\times {10}^{-4}$, but in
panels (b) and (c) the required $D=3\times{10}^{-2}$, i.e. it is 50 times
larger than that in (a). A comparison of results between the closed 
and open systems shows that introducing decoherence 
substantially improves quantum-classical correspondence. Several relevant
observations are in order. First, the quantum resonance in the DB case is
suppressed by increasing decoherence, leading to better agreement between
quantum and classical transmission probabilities. This is discussed
further in Section \ref{resonance}. Second, in order to reach
classicality, the decoherence needed for both the SBS and DB cases is 50
times larger than that of the SBW case, due to the higher nonlinearity of
their potentials. It is tempting to envision that classicality in the
double barrier  case might be reached with smaller decoherence than that
of the single barrier SBS case of the same nonlinearity, since the
quantum DB results shows a resonance that is absent in the SBS case.
This view comes from the semiclassical picture  \cite{miller} suggesting
that the resonance region may  be more sensitive to decoherence than is
the threshold region since resonance require constructive interferences
between trajectories. However, careful examination shows that it is the
nonlinearity that dictates the magnitude of the decoherence required to
reach classicality.

The third observation requires more attention. Note that Fig.
\ref{fig11_0} captures an overall qualitative picture of
decoherence effects on the transmission probabilities. That is, in
both the quantum and classical cases, decoherence enhances the
transmission  at energies below the barrier height and suppresses
it above the barrier height, similar to the trend seen in  \hh2.
This behavior can be understood by considering the increase in
energy width $\delta E$ associated with introducing decoherence of the type
given in Eq. (\ref{operator}). Figure
\ref{fig15} shows the computed growth of  $\delta
E$ for the closed and open systems in the SBS case. The quantum
and classical energy widths are seen to be in good agreement as a
function of time, and become comparable to the 2-unit high
barrier. Noting that the average energy $\langle E\rangle$ for the
open system increases linearly with the rate of $D/m$
\cite{gallis} as well, one can understand from the classical point
of view why decoherence enhances the transmission at $E=1.4$,
suppresses it at $E=3.0$, and has little effect at $E=2.0$ [Fig.
\ref{fig11_0}(b)]. Specifically, at the collision time of $t
\approx 3$, in the presence of decoherence, $\delta E$ is
approximately   1.05, 0.75, and 0.70 [Fig. \ref{fig15}] for the
initial average energy $E$= 3.0, 2.0, and 1.4, respectively, and
is much larger than the average energy increase ($\approx 0.06$).
Consequently, when the wavepacket of $E=3.0$ strikes the barrier, the
decoherence-induced $\delta E$ increases that part of the
wavepacket below the barrier height, increasing the reflected
component, and reducing the transmission probability. (Note that
although the average energy increases, decoherence-induced changes
in the portion of the wavepacket above the barrier
will not alter the transmission, since it will
transmit in any case.) By contrast, for the case of $E=1.4$,
decoherence increases the portion of the energy above the barrier,
allowing it to pass over the barrier, thus increasing  the
transmission probability. For the case of $E=3.0$ and $E=1.4$ the
same trends are seen in both the quantum and classical results. On
the other hand, for $E=2$, equal to the barrier height, the
transmission probability is rather insensitive to the decoherence.
Similar observations apply to the other barrier cases.

\subsection{Resonance Regime}
\label{resonance}

Consider now the resonance region. This region has a characteristic width
denoted $\delta E_R$. It is reasonable to assume that when $D$
is sufficiently large 
to ensure that $\delta E$ is on the
order of $\delta E_R$ at the time of the scattering resonance $t_R$ then
the resonance feature is eliminated. To obtain a crude estimate of the
$D$ for which this occurs consider Eq. (\ref{energywidth}):
\begin{equation}
\frac{d(\delta E)^2}{dt} =2D\langle  p_x^2\rangle/m^2
\approx 4D E_0 /m,
\end{equation}
where $E_0$ is $\langle H \rangle$ at time zero. Here, in order to
obtain a simple estimate, the initial kinetic
energy has been assumed constant in time.
Noting that here $\delta E$ at time zero is ignorable compared to
the growth rate, gives
\begin{equation}
\delta E = \sqrt{4D E_0 t/m}.
\end{equation}
A comparison with Fig. \ref{fig15} shows that this drastic approximation simply
replaces the saturation behavior by linear growth of $\delta E$, overestimating
the growth of $\delta E$.

The simplest $D$ estimate is then that the resonance width $\delta E_R$,
which occurs at time $t_R$, will be washed out for decoherence strengths
$D_R$ on the order of:
\begin{equation}
D_R \approx (\delta E_R)^2 m/[4 E_0 t_R]
\label{approxdr}
\end{equation}

Sample computations using this estimate are shown in Fig.
\ref{testD} for the DB case with varying mass $m$ (which implicitly
changes $\delta E_R$ and $t_R$). In the cases shown,
$\delta E_R$ varied from 0.6 to 0.3 and $D_R$ from 0.007 to
0.0035. In all cases we compare the decoherence free result
(denoted QM) to the result with either decoherence with $D_R$ or,
for comparison, with two different values of $D$, one being $D_R$.
As is evident from Figure \ref{testD} the characteristic resonance
shape disappears with the application of decoherence of magnitude
$D_R$ in each case. Interestingly, as in \hh2, the resonance
region disappears via a lowering of the initial peak that
indicates resonant behavior, leaving a smooth falloff of $P_R$
with decreasing energy.   The smaller values of $D$, shown in
panels (a) and (b) for comparison, are insufficiently large to
completely eliminate the resonance.

\subsection{\hh2}

This argument can be extended to the resonance region in \hh2.
Since our  quantum \hh2 calculation does
not provide the increase of  $\delta E$ (due to absorbing boundary
conditions), $\delta E$ is estimated from the classical
ensemble. At $t=150$ fs $\sim t_f/2$ and at $E=$ 0.9 eV, which is
near the resonance dip, the smaller decoherence magnitude used
above ($D=2.47\times 10^{-35}$ kg$\cdot$J/s) gives an average
energy increase of $\sim$ 0.004 eV, which is relatively small
compared to the average energy, and an energy width $\delta E
\sim$ 0.08 eV, which is roughly comparable to the resonance width.
Hence, one would expect only partial elimination of the resonance, which
is indeed the case [Fig. \ref{fig1m}]. By comparison, the larger
value of $D$ used [Fig. \ref{fig1m2}] is expected to have a
$\delta E$ that is roughly 0.4 eV and an average energy increase
of $\sim$ 0.09 eV, giving the extremely broad featureless results
in Fig. \ref{fig1m2}. The resultant behavior in the \hh2 case, is then
entirely consistent with the broadening due to the decoherence induced
$\delta E$.

\section{Summary}

Decoherence has been shown to significantly alter both
the quantum and classical \hh2 reaction probabilities, leading to
quantum -classical agreement at realistic decoherence values.
Unexpected behavior in the tunneling region, where the reaction
probability is increased below the reaction threshold, but
decreased above the threshold, lead to a more detailed examination
of decoherence in the tunneling regime of simple one dimensional
problems. The results of this study showed that (a) the extent of
potential nonlinearity dictated the extent of decoherence required
to reach quantum-classical agreement: that is, the larger the
nonlinearity, the higher the required decoherence parameter, and
(b) that the increase in energy width, due to environmental-induced 
localization, was responsible for the observed
correspondence in probabilities in the tunneling and resonance
regimes.

Studies of this type also made clear the general dearth of information on
actual decoherence parameters for realistic systems. For this reason,
realistic studies of this type are in progress \cite{yossi2}.

 \vspace{0.2in} {\bf Acknowledgments}: We thank Dr. S. Mahapatra and
Prof. N. Sathyamurthy for sharing a program code for the closed quantum
system in the collinear \hh2 reaction, and Dr. Jiangbin Gong for
discussions. This work was supported by the Natural Sciences and
Engineering Research Council of Canada and by Photonics Research Ontario.

\newpage
\section*{Appendix: Values of $D$}

Within Caldeira-Leggett type models, the constant $D$ is
related to the relaxation constant, $f$ by \cite{phz}
     \begin{eqnarray}
     D=2 f  m kT,
       \end{eqnarray}
where $m$ is the system mass, $k$ the Boltzmann constant, $T$ the
temperature of the environment, and $f$ the strength of the
system-environment coupling.
 Since  a typical  velocity  relaxation time for small molecules
 is  0.01  -  0.1  ps  \cite{rice},
  $D_{typical}$  is  roughly  $10^{-33} -  10^{-34}$ kg$\cdot$J/s 
for the mass of hydrogen at room temperature.
This gives, for our choice of  the smaller value of $D$ that
$D/D_{typical} \sim 10^{-2} - 10^{-1}$.

\newpage
\centerline{{\bf FIGURE CAPTIONS}}

\vspace{0.2in} \noindent
 Figure 1. Collinear reactive transition probabilities in
\hh2 versus the initial total energy $E$, without and with decoherence.
Here $D=2.47\times 10^{-35}$ kg$\cdot$J/s.

\vspace{0.2in} \noindent
Figure 2. As in Fig.\ref{fig1m},  but with 20
times stronger decoherence, \ie, $D=4.94\times 10^{-34}$  kg$\cdot$J/s.

\vspace{0.2in} \noindent
Figure 3. Left panel: Potential  energy of SBW
with scale at left, and the first   derivative   of   SBW  with  scale  at
right.  For comparison, the much sharper SBS potential energy is shown as
well. Right panel: Potential  energies and their first  derivative  for
the  SBS and DB cases. Thin lines and thick lines correspond to SBS and
DB, respectively.

\vspace{0.2in} \noindent Figure 4. Transmission   probabilities as
a function  of initial energy,  $E$,  without  and with
decoherence. (a) SBW case, (b) SBS case, and (c) DB
case. All variables are in dimensionless units.

\vspace{0.2in} \noindent Figure 5. Time dependence  of  $\delta E$
for the open and closed SBS system. Solid  lines  and cross
symbols denote the quantum  and classical results, respectively,
for the open system; from  top to  bottom $E$ takes the values
3.0, 2.0  and  1.4.  A thick dashed line along $\delta E$= 0.1
denotes  the quantum and classical results for the closed system.
All variables are in dimensionless units, and all initial $\delta
E \approx 0.1$.

\vspace{0.2in} \noindent Figure 6. Transmission probabilities for
the DB case, with and without decoherence, for various masses, to
test the predicted $D_R$ [Eq. (\ref{approxdr})]. (a) $m=1$,
$t_R$=6, $\delta E_R$=0.6, $D_R=D_2$=0.006,  $D_1$=0.003, (b)
$m=2$, $t_R$=8, $\delta E_R$=0.5, $D_R=D_2$=0.007, $D_1$=0.002,
(c) $m=3$, $t_R$=9, $\delta E_R$=0.3, $D_R=D$=0.0035.

\newpage
\begin{figure}[htbp]
\centerline{\hbox{\epsfxsize=6.0in \epsfbox{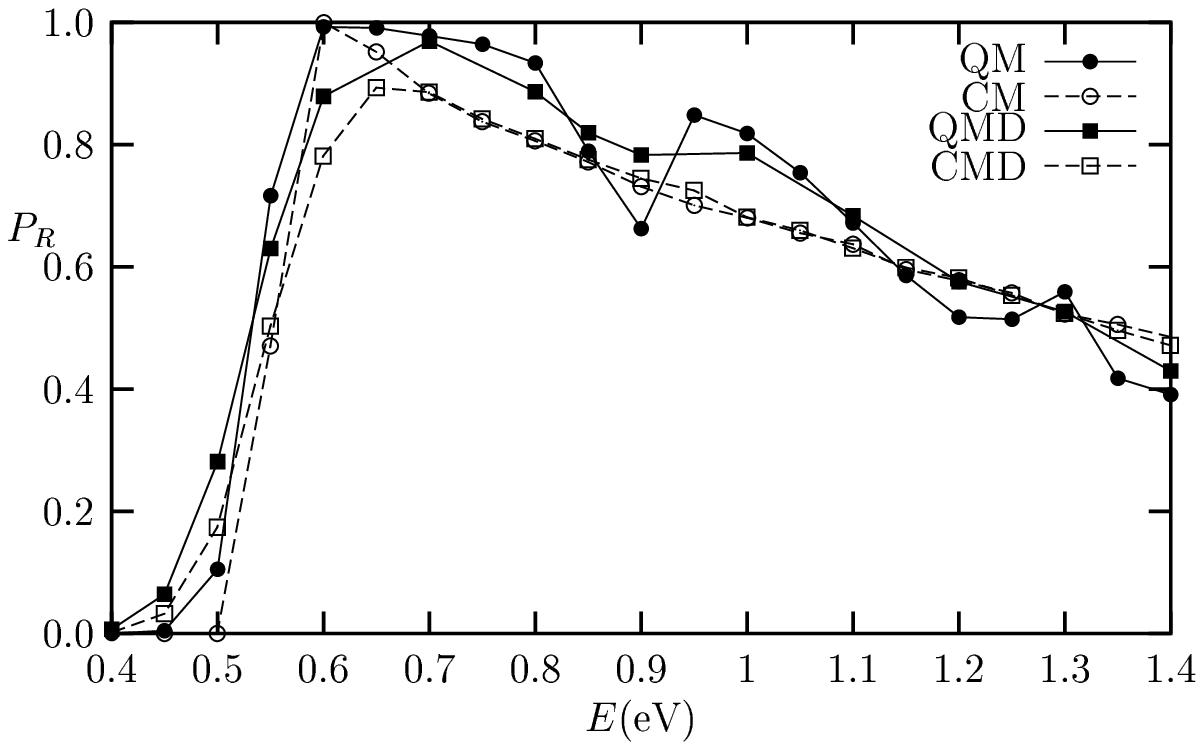}}} \caption{ }
\label{fig1m}
 \end{figure}
\newpage

\begin{figure}[htbp]
\centerline{\hbox{\epsfxsize=6.0in \epsfbox{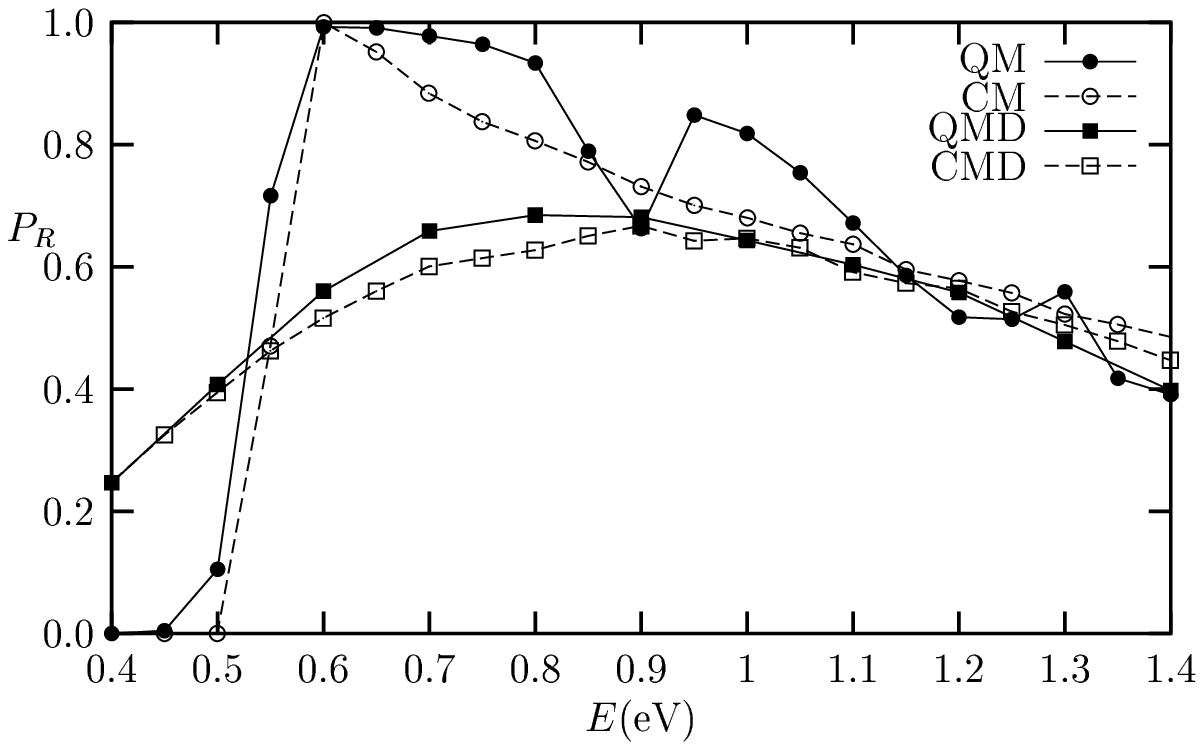}}} \caption{}
 \label{fig1m2}
 \end{figure}
 \newpage

\begin{figure}
\begin{center}
\begin{tabular}{cc}
  {\hbox{\epsfxsize=3.4in
  \epsfbox{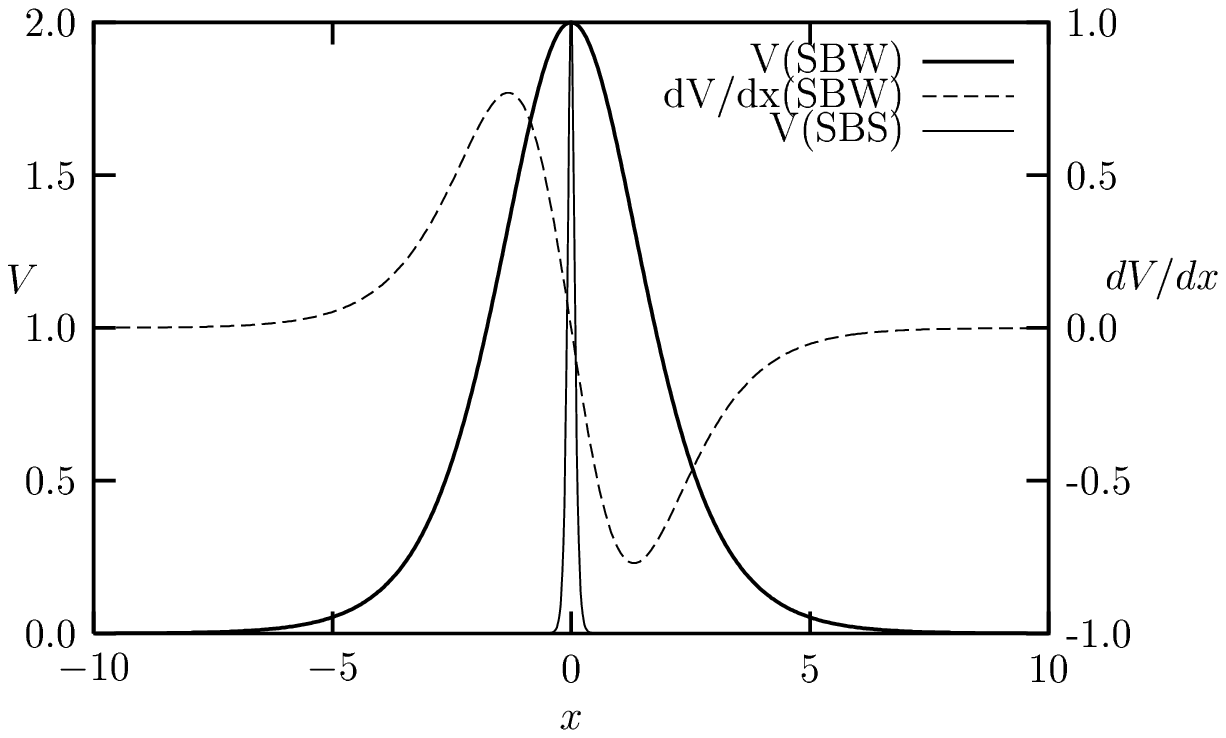}}}{\hbox{\epsfxsize=3.4in
  \epsfbox{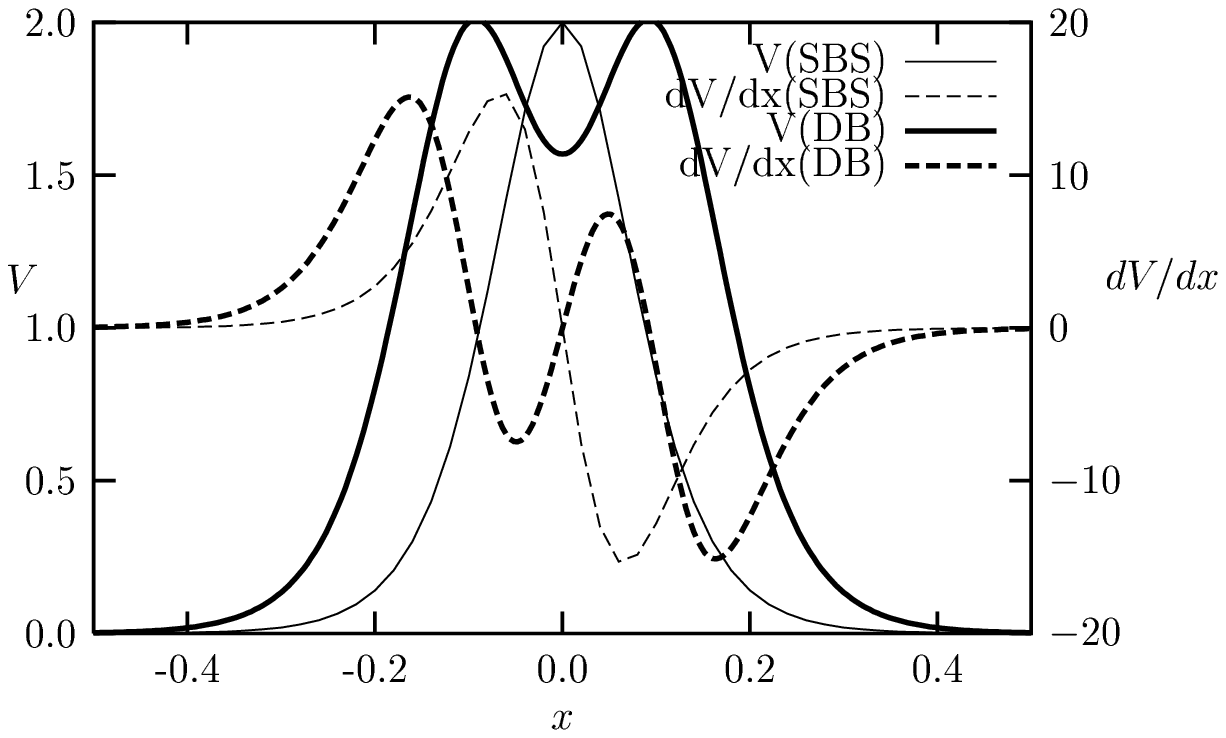}}}\\
  \end{tabular}
   \caption{   }
 \label{fig10}
\end{center}   \end{figure}
\newpage
\begin{figure}
\begin{center}
   \includegraphics[angle=0,scale=.8]{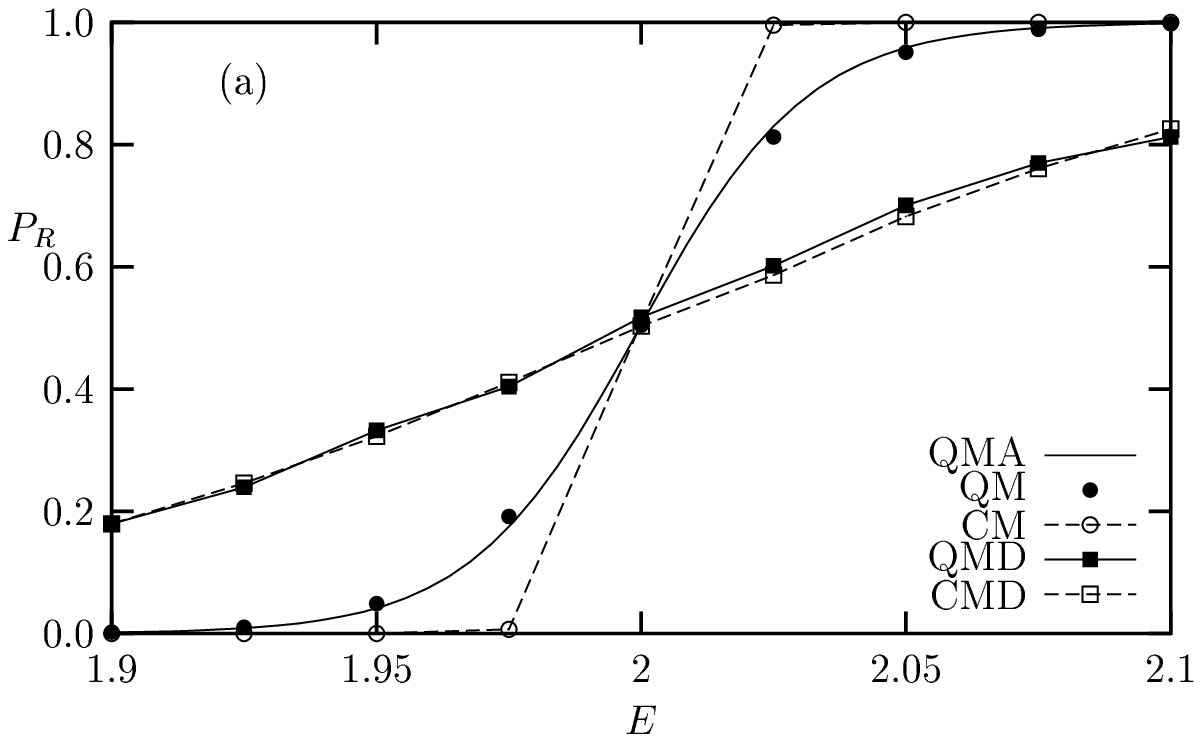}\\
   \includegraphics[angle=0,scale=.8]{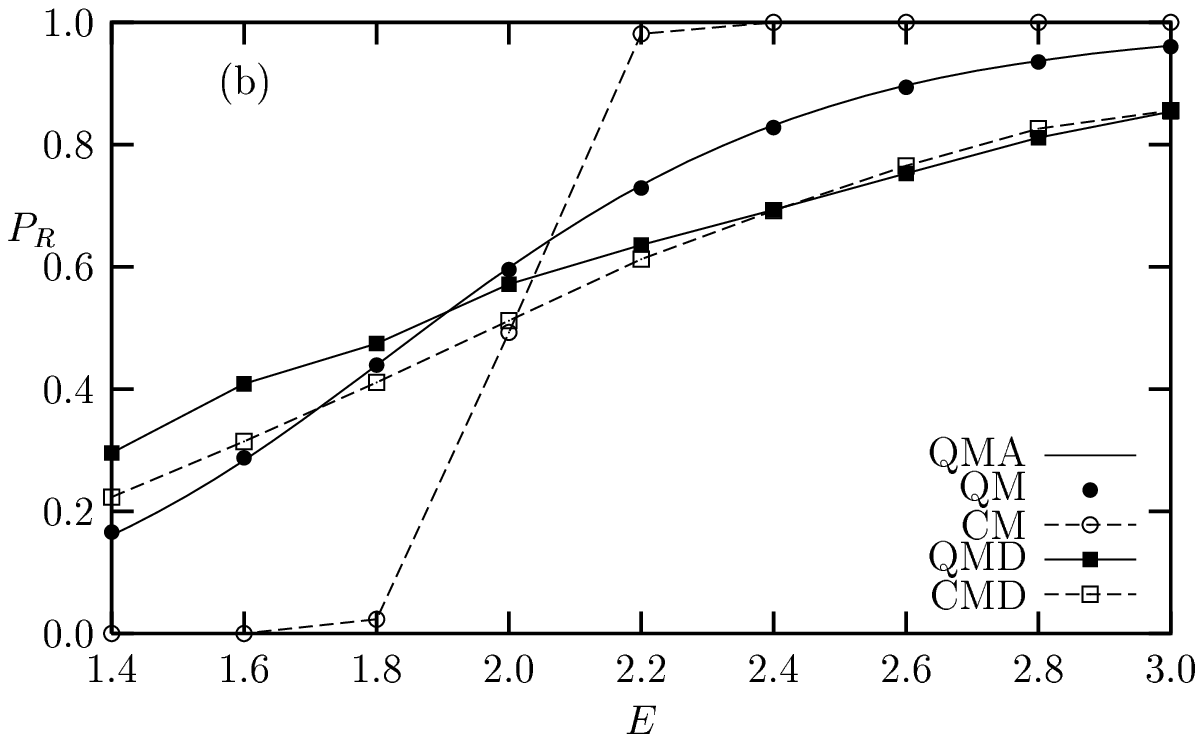}\\
   \includegraphics[angle=0,scale=.8]{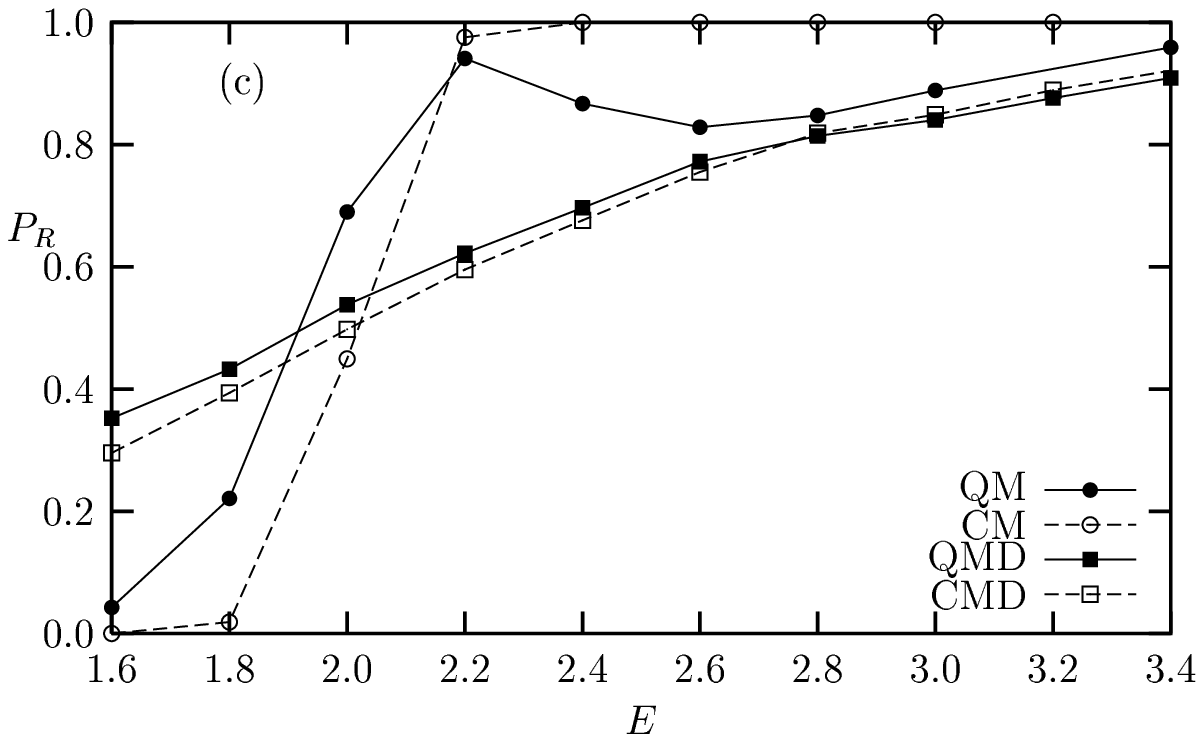}
   \caption{}
 \label{fig11_0}
\end{center}
\end{figure}
\newpage
\begin{figure}
   \centerline{\hbox{\epsfxsize=6.0in \epsfbox{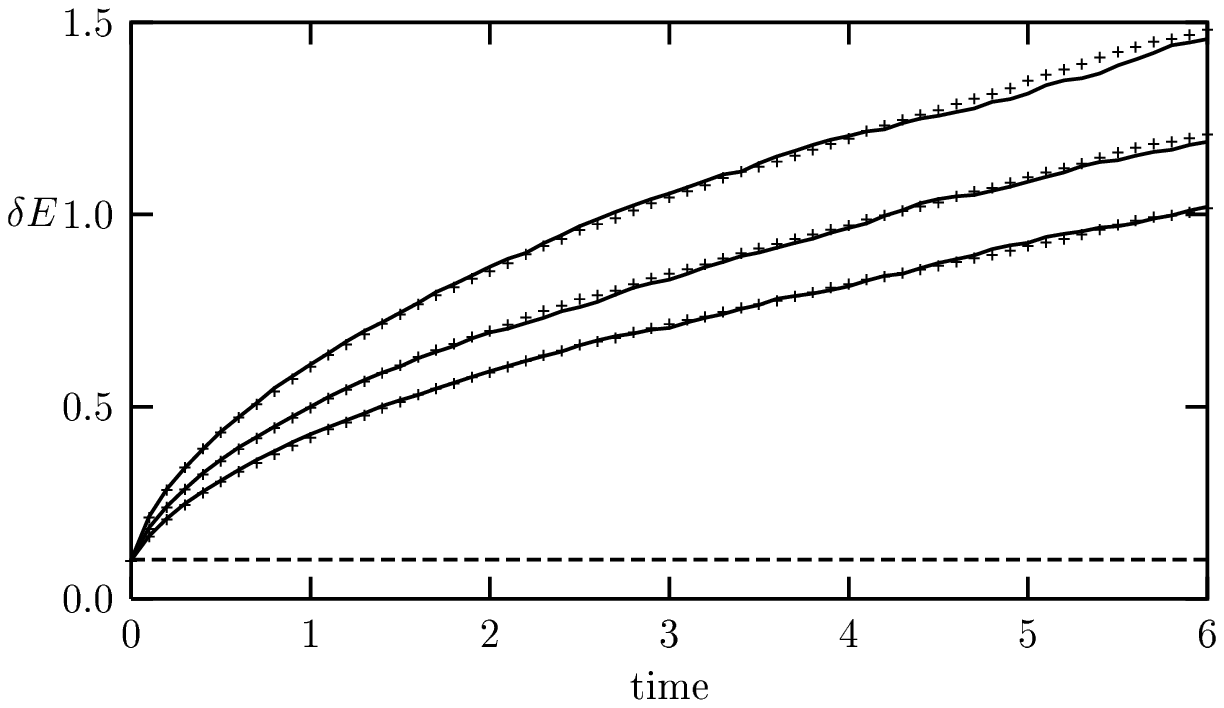}}}
   \caption{ }
   \label{fig15}
   \end{figure}

\begin{figure}

\begin{center}

\includegraphics[angle=0,scale=.8]{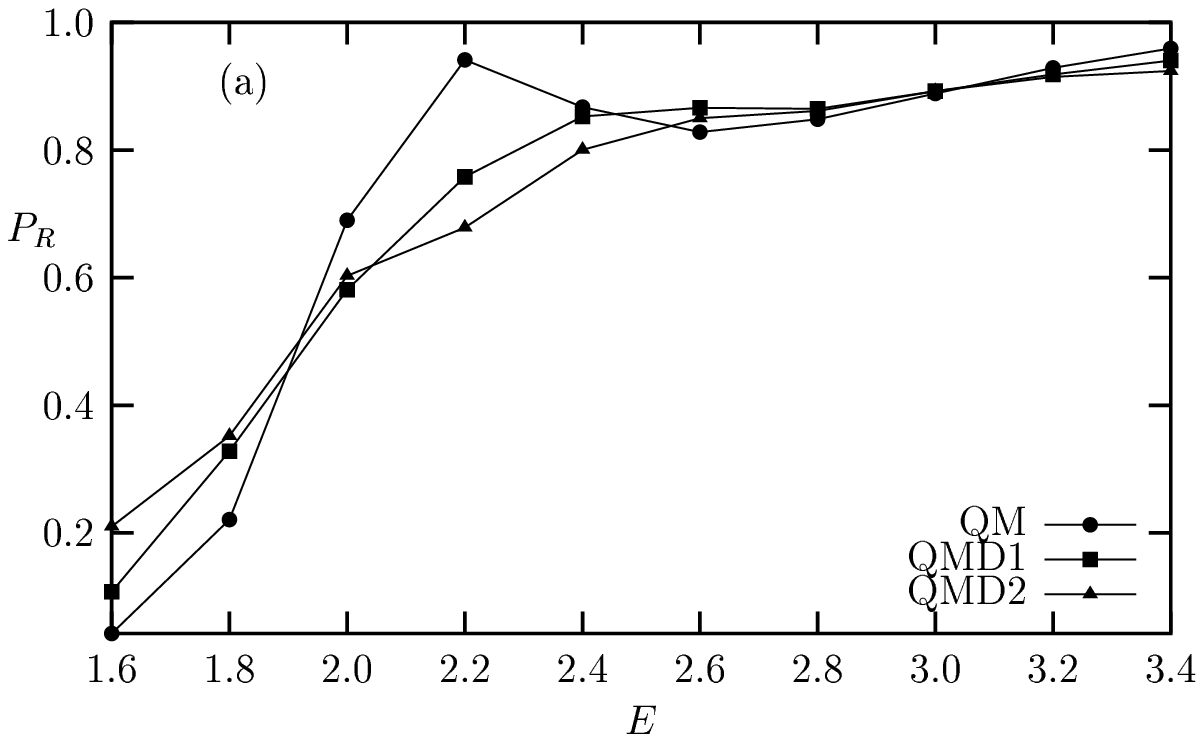}

\includegraphics[angle=0,scale=.8]{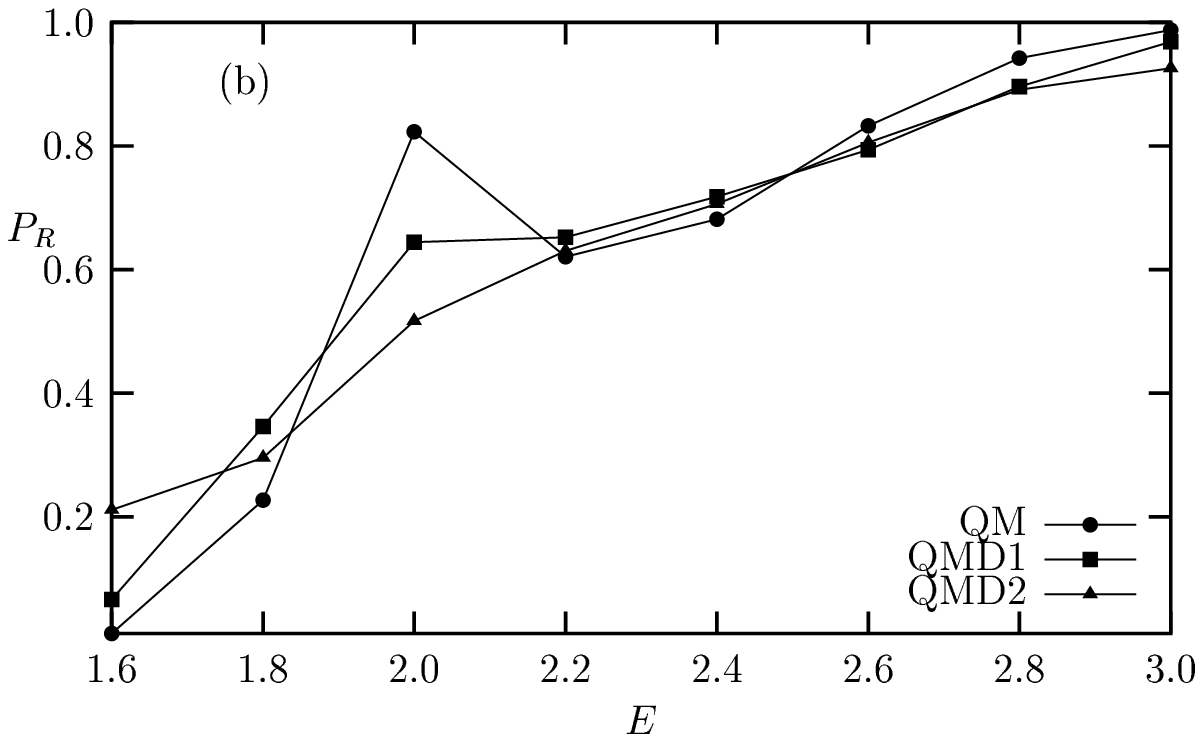}

\includegraphics[angle=0,scale=.8]{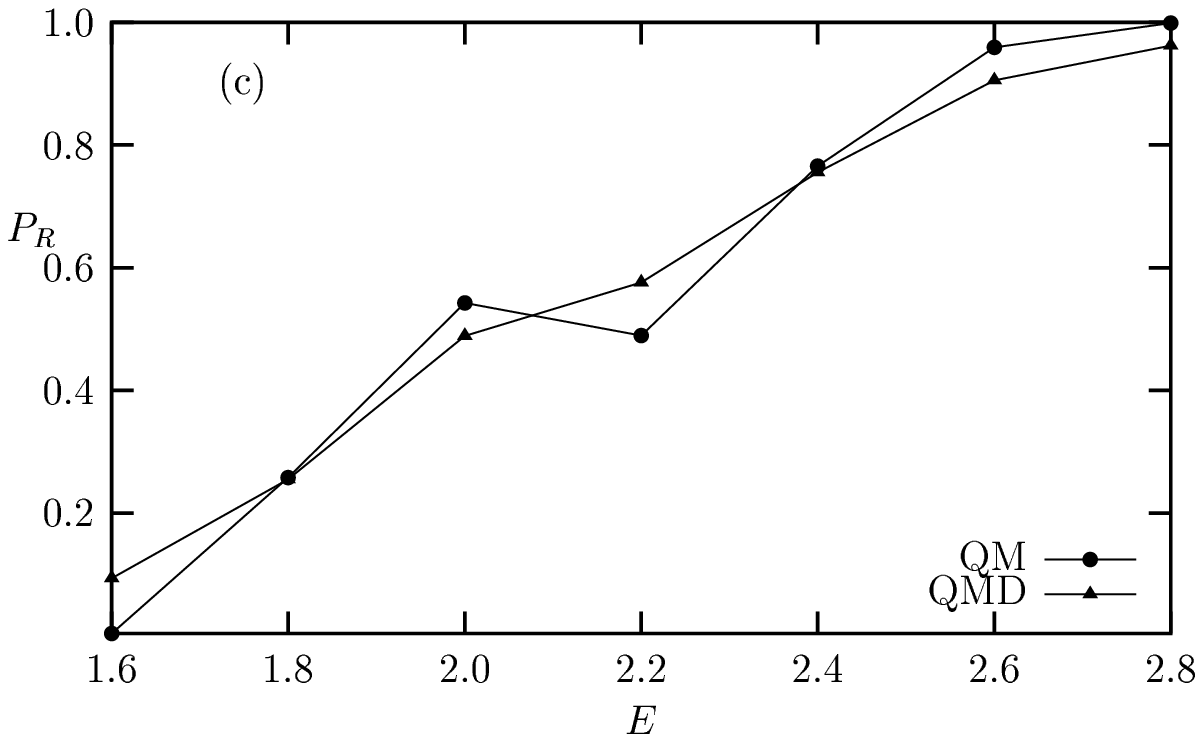}

 \caption{}
 \label{testD}
\end{center}
\end{figure}


\newpage
\begin{thebibliography}{99}

\bibitem{decoherence_review}
 I. O. Stamatescu,   E. Joos, H. D. Zeh, C. Kiefer,  D. Giulini,  and  J. Kupsch, {\it
Decoherence and the Appearance of a Classical World in Quantum
Theory}, 2nd ed. (Springer Verlag, 2003); W. H. Zurek, Rev. Mod.
Phys. {\bf 75}, 715 (2003); P. Blanchard, D. Giullini, E. Joos,
C. Kiefer, I. O. Stamatescu (eds), {\it Decoherence: Theoretical,
Experimental, and Conceptual Problems}, 2nd ed. (Lecture Notes in
Physics 538, Springer Verlag, 2000).


\bibitem{joos} E. Joos and H.D. Zeh Z. Phys. B 59, 223 (1985)
\bibitem{DPE}
A.O. Caldeira and A.J. Leggett, Physica {\bf 121A}, 587 (1983).

\bibitem{unruh}
W.G. Unruh and W.H. Zurek, Phys. Rev. D {\bf 40}, 1071 (1989); B.L. Hu,
J.P. Paz and Y.  Zhang, Phys. Rev.D {\bf 45},
 2843 (1992); {\bf 47}, 1576 (1993)

\bibitem{hh2research}
R.  Sadeghi  and  R.  T. Skodje, J. Chem. Phys. {\bf 99}, 5126 (1993) and
references are therein.



\bibitem{Wilkie} J. Wilkie and P. Brumer. Phys. Rev. A 55, 27 (1997);
{\em ibid.} Phys. Rev. A 55, 43 (1997).

\bibitem{zurek}
S. Habib, K. Shizume, and W. H. Zurek, \prl {\bf 80}, 4361 (1998)
and references are therein.
\bibitem{gong} J. Gong and P. Brumer, Phys. Rev. E {\bf 60}, 1643 (1999).


\bibitem{rossky}
E. Bittner and P. Rossky, J. Chem. Phys. {\bf 103}, 8130 (1995).
\bibitem{prezhdo}
O. Prezhdo, Phys. Rev. Lett. {\bf 85}, 4413 (2000).
\bibitem{brumer_shapiro}
M. Shapiro and P. Brumer, Adv. At. Mol. Opt. Phys. {\bf 42}, 287
(2000).
\bibitem{gallis} M. R. Gallis and G.N. Fleming, Phys. Rev. A 42, 38 (1990);
ibid. A 43, 5778 (1991).

\bibitem{perci1}
N. Gisin and I. Percival, J. Phys. A {\bf 25} 5677 (1992) and
references are therein.
\bibitem{fleck}  M.D. Feit, J. A. Fleck and A. Steiger, J. Comput. Phys.
{\bf 47}, 412 (1982).
\bibitem{secondorder}
J. R. Klauder and W. P. Petersen, SIAM J. Numer. Anal. {\bf 22},
1153 (1985).
\bibitem{LSTH}
B. Liu, J. Chem. Phys. {\bf 58}, 1925 (1973); P. Siegbahn and B.
Liu, J. Chem. Phys. {\bf 68}, 2457 (1978); D. G. Truhlar and C. J.
Horowitz, J. Chem. Phys. {\bf 68}, 2466 (1978).
\bibitem{wyatt}
E. A. McCullough, Jr. and R. E. Wyatt, J. Chem. Phys. {\bf 54},
3578 (1971).
\bibitem{kuppermann}
J. M. Bowman and A. Kuppermann, J. Chem. Phys. {\bf 59}, 6524
(1973) and references therein.
\bibitem{wignercm}
H. Lee and T. George,  J. Chem. Phys. {\bf 84}, 6247 (1986).
\bibitem{millerr}
C. C. Rankin and W. H. Miller, J. Chem. Phys. {\bf 55}, 3150
(1971).
\bibitem{yossi} Y. Elran and P. Brumer, J. Chem. Phys. 
{\bf 121}, 2673 (2004); Y. Elran, R. Kapral and P. Brumer, (manuscript
in preparation).
\bibitem{thesis}
H. Han, Ph. D. Dissertation, University of Toronto, (2004).

\bibitem{maskf}
S. Mahapatra and N. Sathyamurthy, J. Chem. Phys.
 {\bf 105}, 10934 (1996).

\bibitem{threshold}
H. Ushiyama and K. Takatsuka, J. Chem. Phys. {\bf 109}, 9664 (1998).

\bibitem{model}
W.H. Zurek and J. P. Paz, Phys. Rev. Lett. {\bf 72}, 2508 (1994);
A. R. Kolovsky, Phys. Rev. Lett. {\bf 76}, 340 (1996); A. R.
Kolovsky, Europhys. Lett. {\bf 27}, 79 (1994).

\bibitem{numbers} Here $\Delta t=0.002$ for the closed quantum systems in all
cases. For SBW, $\Delta t= 0.002/4$ for QMD. For the SBS and DB cases of
the higher nonlinearity, $\Delta t= 0.002/16$ for QMD. For the
grid size of quantum calculation, $\Delta x = 0.068$ for SBW and $\Delta x
= 0.02$ for SBS and DB. The quantity $x_0$ is chosen so that the initial
wavepacket is placed in the asymptotic (force-free) region. In addition,
$\gamma$ is chosen carefully so that for SBW $\delta E \leq 0.01$ and for
SBS and DB $\delta E \leq 0.1$, in other words, so that the wavefunctions
belonging to different initial energies should not overlap one another
since the energy interval of interest is taken as 0.025 for SBW and as 0.2
for SBS and DB.
\bibitem{liboff}
R. Liboff,  {\it Quantum Mechanics}, 2nd ed. (Addison-Wesley,
Massachusetts, 1992).
\bibitem{miller}
For  a  thorough  discussion  of  the semiclassical $S$-matrix
approach  in  molecular  collisions,  see W. H. Miller, Adv. Chem.
Phys. {\bf 30}, 77 (1975).
\bibitem{yossi2}
Y. Elran and P. Brumer (work in progress)
\bibitem{phz}
J. P.  Paz,  S. Habib, and W.H. Zurek, Phys. Rev. D {\bf 47}, 488 (1993)
\bibitem{rice} S. A. Rice, {\it Diffusion-limited  Reactions}  (Elsevier, Amsterdam,
1985).


\end{thebibliography}
\end{document}